\documentclass[submission,copyright,creativecommons]{eptcs}
\providecommand{\event}{QPL 2024} 

\usepackage{iftex}
\usepackage{braket}
\usepackage{tikzit}
\usepackage{amsfonts,amssymb}
\usepackage{graphicx,subcaption}

\ifpdf
  \usepackage{underscore}         
  \usepackage[T1]{fontenc}        
\else
  \usepackage{breakurl}           
\fi

\tikzstyle{gate}=[shape=rectangle, text height=1.5ex, text depth=0.25ex, yshift=0.5mm, fill=white, draw=black, minimum height=5mm, yshift=-0.5mm, minimum width=5mm, font={\small}, tikzit category=circuit]
\tikzstyle{big gate}=[shape=rectangle, text height=1.5ex, text depth=0.25ex, yshift=0.5mm, fill=white, draw=black, minimum height=10mm, yshift=-0.5mm, minimum width=5mm, font={\small}, tikzit category=circuit]
\tikzstyle{Z dot}=[inner sep=0mm, minimum size=2mm, shape=circle, draw=black, fill={rgb,255: red,221; green,255; blue,221}, tikzit category=zx]
\tikzstyle{Z phase dot}=[minimum size=5mm, font={\footnotesize\boldmath}, shape=rectangle, rounded corners=2mm, inner sep=0.2mm, outer sep=-2mm, scale=0.8, tikzit shape=rectangle, draw=black, fill={rgb,255: red,221; green,255; blue,221}, tikzit draw=blue, tikzit category=zx]
\tikzstyle{X dot}=[Z dot, shape=circle, draw=black, fill={rgb,255: red,255; green,136; blue,136}, tikzit category=zx]
\tikzstyle{X phase dot}=[Z phase dot, tikzit shape=rectangle, tikzit draw=blue, fill={rgb,255: red,255; green,136; blue,136}, font={\footnotesize\boldmath}, tikzit category=zx]
\tikzstyle{hadamard}=[fill=yellow, draw=black, shape=rectangle, inner sep=0.6mm, minimum height=1.5mm, minimum width=1.5mm, tikzit category=zx]
\tikzstyle{paulibox}=[fill={rgb,255: red,221; green,221; blue,255}, draw=black, shape=rectangle, inner sep=0.6mm, minimum height=5mm, minimum width=5mm, font={\footnotesize}, text height=1.5ex, text depth=0.25ex, tikzit category=zx]
\tikzstyle{vertex}=[inner sep=0mm, minimum size=1mm, shape=circle, draw=black, fill=black, tikzit category=misc]
\tikzstyle{vertex set}=[inner sep=0mm, minimum size=1mm, shape=circle, draw=black, fill=white, font={\footnotesize\boldmath}, tikzit category=misc]
\tikzstyle{small black dot}=[fill=black, draw=black, shape=circle, inner sep=0pt, minimum width=1.2mm, tikzit category=circuit]
\tikzstyle{cnot ctrl}=[fill=black, draw=black, shape=circle, inner sep=0pt, minimum width=1.2mm, tikzit category=circuit]
\tikzstyle{cnot targ}=[fill=white, draw=white, shape=circle, tikzit category=circuit, label={center:$\oplus$}, inner sep=0pt, minimum width=2.1mm, tikzit fill={rgb,255: red,102; green,204; blue,255}, tikzit draw=black]
\tikzstyle{ket}=[fill=white, draw=black, shape=regular polygon, regular polygon sides=3, regular polygon rotate=-30, scale=0.7, inner sep=1pt, tikzit category=circuit, tikzit shape=rectangle, tikzit fill=green]
\tikzstyle{bra}=[fill=white, draw=black, shape=regular polygon, regular polygon sides=3, regular polygon rotate=30, scale=0.7, inner sep=1pt, tikzit category=circuit, tikzit shape=rectangle, tikzit fill=red]
\tikzstyle{scalar}=[shape=rectangle, text height=1.5ex, text depth=0.25ex, yshift=0.5mm, fill=white, draw=black, minimum height=5mm, yshift=-0.5mm, minimum width=5mm, font={\small}]
\tikzstyle{clabel}=[fill=white, draw=none, shape=rectangle, tikzit fill={rgb,255: red,56; green,255; blue,242}, font={\footnotesize}, inner sep=1pt, tikzit category=labels]
\tikzstyle{empty diagram}=[draw={gray!40!white}, dashed, shape=rectangle, minimum width=1cm, minimum height=1cm, tikzit category=misc]
\tikzstyle{amap}=[fill=white, draw=black, shape=NEbox, tikzit category=asymmetric, tikzit fill=yellow, tikzit shape=rectangle]
\tikzstyle{amap conj}=[fill=white, draw=black, shape=NWbox, tikzit category=asymmetric, tikzit fill=green, tikzit shape=rectangle]
\tikzstyle{amap adj}=[fill=white, draw=black, shape=SEbox, tikzit category=asymmetric, tikzit fill=red, tikzit shape=rectangle]
\tikzstyle{amap trans}=[fill=white, draw=black, shape=SWbox, tikzit category=asymmetric, tikzit fill=orange, tikzit shape=rectangle]
\tikzstyle{astate}=[fill=white, draw=black, shape=NEtriangle, tikzit category=asymmetric, tikzit shape=circle, tikzit fill=yellow]
\tikzstyle{astate conj}=[fill=white, draw=black, shape=NWtriangle, tikzit category=asymmetric, tikzit shape=circle, tikzit fill=green]
\tikzstyle{astate adj}=[fill=white, draw=black, shape=SEtriangle, tikzit category=asymmetric, tikzit shape=circle, tikzit fill=red]
\tikzstyle{astate trans}=[fill=white, draw=black, shape=SWtriangle, tikzit category=asymmetric, tikzit shape=circle, tikzit fill=orange]

\tikzstyle{hadamard edge}=[-, dashed, dash pattern=on 2pt off 0.5pt, thick, draw={rgb,255: red,68; green,136; blue,255}]
\tikzstyle{box edge}=[-, dashed, dash pattern=on 2pt off 0.5pt, thick, draw={rgb,255: red,203; green,192; blue,225}]
\tikzstyle{brace edge}=[-, tikzit draw=blue, decorate, decoration={brace,amplitude=1mm,raise=-1mm}]
\tikzstyle{diredge}=[->]
\tikzstyle{double edge}=[-, double, shorten <=-1mm, shorten >=-1mm, double distance=2pt]
\tikzstyle{gray edge}=[-, {gray!60!white}]
\tikzstyle{pointer edge}=[->, very thick, gray]
\tikzstyle{boldedge}=[-, line width=1.6pt, shorten <=-0.17mm, shorten >=-0.17mm]
\tikzstyle{bidir edge}=[<->, very thick, draw={rgb,255: red,191; green,191; blue,191}]
\tikzstyle{separator edge}=[-, dashed, dash pattern=on 2pt off 0.5pt, thick, draw={rgb,255: red,153; green,153; blue,153}]

\title{Procedurally Optimised ZX-Diagram Cutting for Efficient T-Decomposition in Classical Simulation}
\author{Matthew Sutcliffe
\institute{Department of Computer Science\\ University of Oxford\\ Oxford, UK}
\email{matthew.sutcliffe@cs.ox.ac.uk}
\and
Aleks Kissinger
\institute{Department of Computer Science\\ University of Oxford\\ Oxford, UK}
\email{aleks.kissinger@cs.ox.ac.uk}
}
\def\titlerunning{Procedurally Optimised ZX-diagram Cutting}
\def\authorrunning{M. Sutcliffe \& A. Kissinger}
\begin{document}
\maketitle

\begin{abstract}
A quantum circuit may be strongly classically simulated with the aid of ZX-calculus by decomposing its $t$ T-gates into a sum of $2^{\alpha t}$ classically computable stabiliser terms. In this paper, we introduce a general procedure to find an optimal pattern of vertex cuts in a ZX-diagram to maximise its T-count reduction at the cost of the fewest cuts. Rather than reducing a Clifford+T diagram based on a fixed routine of decomposing its T-gates directly (as is the conventional approach), we focus instead on taking advantage of certain patterns and structures common to such circuits to, in effect, design by automatic procedure an arrangement of spider decompositions that is optimised for the particular circuit. In short, this works by assigning weights to vertices based on how many T-like gates they are blocking from fusing/cancelling and then appropriately propagating these weights through any neighbours which are then blocking weighted vertices from fusing, and so on. Ultimately, this then provides a set of weightings on relevant nodes, which can then each be cut, starting from the highest weighted down. While this is a heuristic approach, we show that, for circuits small enough to verify, this method achieves the most optimal set of cuts possible $71\%$ of the time. Furthermore, there is no upper bound for the efficiency achieved by this method, allowing, in principle, an effective decomposition efficiency $\alpha\rightarrow0$ for highly structured circuits. Even applied to random pseudo-structured circuits (produced from CNOTs, phase gates, and Toffolis), we record the number of stabiliser terms required to reduce all T-gates, via our method as compared to that of the more conventional T-decomposition approaches (namely \cite{kissinger21}, with $\alpha\approx0.47$), and show consistent improvements of orders of magnitude, with an effective efficiency $0.1\lesssim\alpha\lesssim0.2$.
\end{abstract}

\section{Introduction}

Present-day quantum hardware is very limited, with few qubits and much noise \cite{preskill18}. Consequently, there are many classical techniques that are often employed to better optimise quantum circuits and/or to verify the behaviour of quantum hardware and software. A particularly useful tool for facilitating this is ZX-calculus \cite{CD2,wetering20,Backens17,pqp,pqs}, which allows quantum circuits to be expressed and simplified graphically with the use of known rewriting rules. This has been utilised for various problems in the field, including notably optimisation \cite{duncan2019graph,Cowtan2020phasegadget,deBeaudrapN2020treducspidernest,borgna2021hybrid,gogioso2023annealing,mcelvanney2023flowpreserving,nagele2023optimizing} and classical simulation \cite{kissinger21,kissinger2022classical,10.1145/3489517.3530627,Codsi2022Masters,codsi2022classically,cam2023speeding,koch2023speedy}. On the latter problem, for instance, wherein one wishes to compute the probabilities of particular measurement outcomes of a quantum circuit, ZX-calculus can be utilised to re-express a large `Clifford+T' circuit (which is notoriously inefficient to simulate classically) as a sum of `Clifford' circuits (which \textit{are} efficient to classically simulate) \cite{kissinger21}.

Re-expressing a Clifford+T circuit as sum of Clifford circuits in this way relies upon known decompositions of sets of costly `T-gates' into cheap `Clifford' terms. There are many such decompositions that have been discovered \cite{BSS,bravyi19,kissinger21,kissinger2022classical}, of varying efficiencies, and the conventional strategies \cite{kissinger21}, perhaps unsurprisingly, tend to opt for the most efficient (i.e. those which translate a set of T-gates into the fewest number of Clifford terms). However, the apparent efficiencies of these decompositions can be misleading. In fact, it has been noted that applying apparently \textit{less} efficient decompositions in certain circumstances can result in \textit{more} efficient results overall (that is, fewer Clifford terms in exchange for removing all T-gates of a circuit) \cite{Codsi2022Masters}. Knowing when this can be applicable on local scales is often fairly straightforward. However, what is seldom considered is how appropriate applications of these less efficient decompositions can be found when looking on broader scales.

Specifically, in this paper, we demonstrate how drastic improvements to the overall efficiency can be attained by applying such decompositions to inconspicuous gates in a circuit with no obvious or immediate local benefit for doing so. We then present a means by which one can procedurally analyse a given circuit to determine where these optimal gates for decomposition are. We do this based on a heuristic of weighing vertices due to their immediate local benefit of applying such a decomposition, and then propagating these weights through their neighbours as appropriate. Lastly, we showcase the effectiveness of our method by comparing its overall efficiency at fully decomposing various Clifford+T circuits, of various numbers of T-gates, against the more traditional methods \cite{kissinger21}.

\section{Background}

\subsection{ZX-calculus}
\label{subsec:zxcalc}

A very useful notation with which to express quantum circuits is that provided by \textit{ZX-calculus} \cite{CD2,wetering20,Backens17}, wherein all operations are expressed as \textit{spiders} (phase rotations) about either the Z-axis or X-axis of the Bloch sphere, connected via \textit{edges} (or \textit{wires}). Circuits expressed in this notation are known as \textit{ZX-diagrams} and, by convention, Z-spiders are green and X-spiders red - in either case with the angle of rotation written within (or left blank if zero), as follows:

\ctikzfig{spider_def}

In addition to these two types of spiders, for convenience the Hadamard gate is often also included explicitly, as a yellow box, though this too may be decomposed (via the Euler decomposition) into spiders as such:

\ctikzfig{had_def}

As shown here, an edge containing a Hadamard (referred to, unsurprisingly, as a \textit{Hadamard edge}) may alternatively be expressed as a dashed blue line.

From these basic components, any quantum circuit may be expressed as a ZX-diagram. Indeed, completeness for the Clifford gateset is achieved provided phases of $\frac{n\pi}{2}$ are allowed (where $n\in\mathbb{Z}$). (Note also that, as phases correspond to rotations, they are all of modulo $2\pi$.) To extend the completeness to the Clifford+T gateset (including T-gates and Toffoli gates), the resolution must be expanded to support phases of $\frac{n\pi}{4}$.

\subsection{ZX-diagram rewriting}

\begin{figure}[h]
\begin{tikzpicture}
	\begin{pgfonlayer}{nodelayer}
		\node [style=none] (0) at (-2, 1) {};
		\node [style=none] (1) at (-2, -0.25) {};
		\node [style=Z phase dot] (4) at (-0.75, 0.25) {$\alpha$};
		\node [style=none] (6) at (0.5, 1) {};
		\node [style=none] (8) at (0.25, 0.5) {$\vdots$};
		\node [style=none] (9) at (-1.75, 0.5) {$\vdots$};
		\node [style=none] (10) at (1.5, -0.75) {$=$};
		\node [style=none] (12) at (-2, 1.5) {};
		\node [style=none] (13) at (0.5, 1.5) {};
		\node [style=none] (14) at (0.5, -0.25) {};
		\node [style=none] (15) at (-2, -2.5) {};
		\node [style=none] (16) at (-2, -3) {};
		\node [style=Z phase dot] (17) at (-0.75, -1.75) {$\beta$};
		\node [style=none] (18) at (0.5, -2.5) {};
		\node [style=none] (19) at (0.25, -1.75) {$\vdots$};
		\node [style=none] (20) at (-1.75, -1.75) {$\vdots$};
		\node [style=none] (21) at (-2, -1.25) {};
		\node [style=none] (22) at (0.5, -1.25) {};
		\node [style=none] (23) at (0.5, -3) {};
		\node [style=none] (24) at (-0.75, -0.75) {$...$};
		\node [style=none] (25) at (2.5, 0.75) {};
		\node [style=Z phase dot] (27) at (3.75, -0.75) {$\;\alpha+\beta\;$};
		\node [style=none] (28) at (5, 0.75) {};
		\node [style=none] (31) at (2.5, 1.5) {};
		\node [style=none] (32) at (5, 1.5) {};
		\node [style=none] (38) at (4.75, -0.5) {$\vdots$};
		\node [style=none] (39) at (2.75, -0.5) {$\vdots$};
		\node [style=none] (40) at (2.5, -3) {};
		\node [style=none] (41) at (5, -3) {};
		\node [style=none] (42) at (2.5, -2.25) {};
		\node [style=none] (43) at (5, -2.25) {};
		\node [style=Z phase dot] (46) at (-1.25, -6.25) {$\alpha$};
		\node [style=none] (47) at (0, -5.25) {};
		\node [style=none] (48) at (-0.25, -6.25) {$\vdots$};
		\node [style=none] (50) at (-2.5, -6.25) {};
		\node [style=none] (51) at (0, -5.75) {};
		\node [style=none] (52) at (0, -7.25) {};
		\node [style=hadamard] (53) at (-2, -6.25) {};
		\node [style=none] (54) at (0.5, -5.25) {};
		\node [style=none] (55) at (0.5, -5.75) {};
		\node [style=none] (56) at (0.5, -7.25) {};
		\node [style=none] (57) at (1.5, -6.25) {$=$};
		\node [style=X phase dot] (58) at (4, -6.25) {$\alpha$};
		\node [style=hadamard] (59) at (5.25, -5.25) {};
		\node [style=none] (60) at (5, -6.25) {$\vdots$};
		\node [style=none] (61) at (2.75, -6.25) {};
		\node [style=hadamard] (62) at (5.25, -5.75) {};
		\node [style=hadamard] (63) at (5.25, -7.25) {};
		\node [style=none] (64) at (3.25, -6.25) {};
		\node [style=none] (65) at (5.75, -5.25) {};
		\node [style=none] (66) at (5.75, -5.75) {};
		\node [style=none] (67) at (5.75, -7.25) {};
		\node [style=none] (78) at (12.25, -0.75) {$=$};
		\node [style=Z phase dot] (79) at (14.75, -0.75) {$\;(-1)^{a}\alpha\;$};
		\node [style=X phase dot] (80) at (16.25, 1) {$a\pi$};
		\node [style=none] (81) at (16.25, -1) {$\vdots$};
		\node [style=none] (82) at (13.25, -0.75) {};
		\node [style=X phase dot] (83) at (16.25, 0) {$a\pi$};
		\node [style=X phase dot] (84) at (16.25, -2.25) {$a\pi$};
		\node [style=none] (85) at (14, -0.75) {};
		\node [style=none] (86) at (17, 1) {};
		\node [style=none] (87) at (17, 0) {};
		\node [style=none] (88) at (17, -2.25) {};
		\node [style=none] (89) at (14, -1.75) {$e^{ia\alpha}$};
		\node [style=Z phase dot] (90) at (9, -0.75) {$\alpha$};
		\node [style=none] (91) at (10.5, 1) {};
		\node [style=none] (92) at (10.5, -1) {$\vdots$};
		\node [style=none] (93) at (7.25, -0.75) {};
		\node [style=none] (94) at (10.5, 0) {};
		\node [style=none] (95) at (10.5, -2.25) {};
		\node [style=X phase dot] (96) at (8, -0.75) {$a\pi$};
		\node [style=none] (97) at (11.25, 1) {};
		\node [style=none] (98) at (11.25, 0) {};
		\node [style=none] (99) at (11.25, -2.25) {};
		\node [style=none] (100) at (23.75, -0.75) {$=$};
		\node [style=X phase dot] (102) at (26.75, 1) {$a\pi$};
		\node [style=none] (103) at (26.75, -1) {$\vdots$};
		\node [style=X phase dot] (105) at (26.75, 0) {$a\pi$};
		\node [style=X phase dot] (106) at (26.75, -2.25) {$a\pi$};
		\node [style=none] (108) at (27.5, 1) {};
		\node [style=none] (109) at (27.5, 0) {};
		\node [style=none] (110) at (27.5, -2.25) {};
		\node [style=none] (111) at (25.25, -0.75) {$\frac{e^{ia\alpha}}{\sqrt{2}^{(n-1)}}$};
		\node [style=Z phase dot] (112) at (20.5, -0.75) {$\alpha$};
		\node [style=none] (113) at (22, 1) {};
		\node [style=none] (114) at (22, -1) {$\vdots$};
		\node [style=none] (116) at (22, 0) {};
		\node [style=none] (117) at (22, -2.25) {};
		\node [style=X phase dot] (118) at (19.5, -0.75) {$a\pi$};
		\node [style=none] (119) at (22.75, 1) {};
		\node [style=none] (120) at (22.75, 0) {};
		\node [style=none] (121) at (22.75, -2.25) {};
		\node [style=none] (122) at (8.5, -5.5) {};
		\node [style=none] (123) at (8.5, -7) {};
		\node [style=X dot] (124) at (9.25, -6.25) {};
		\node [style=Z dot] (125) at (10.5, -6.25) {};
		\node [style=none] (126) at (11.25, -5.5) {};
		\node [style=none] (127) at (11.25, -7) {};
		\node [style=none] (128) at (12.25, -6.25) {$=$};
		\node [style=Z dot] (129) at (15.25, -5.5) {};
		\node [style=X dot] (130) at (16.5, -5.5) {};
		\node [style=none] (131) at (14.5, -5.5) {};
		\node [style=none] (132) at (17.25, -5.5) {};
		\node [style=Z dot] (133) at (15.25, -7) {};
		\node [style=X dot] (134) at (16.5, -7) {};
		\node [style=none] (135) at (14.5, -7) {};
		\node [style=none] (136) at (17.25, -7) {};
		\node [style=none] (137) at (13.5, -6.25) {$\sqrt{2}$};
		\node [style=none] (138) at (20.25, -5.5) {};
		\node [style=none] (139) at (22.75, -5.5) {};
		\node [style=Z dot] (140) at (21.5, -5.5) {};
		\node [style=none] (141) at (23.75, -5.5) {$=$};
		\node [style=none] (144) at (20.25, -7) {};
		\node [style=none] (145) at (22.75, -7) {};
		\node [style=none] (147) at (23.75, -7) {$=$};
		\node [style=hadamard] (150) at (21, -7) {};
		\node [style=hadamard] (151) at (22, -7) {};
		\node [style=none] (152) at (24.75, -5.5) {};
		\node [style=none] (153) at (27.25, -5.5) {};
		\node [style=none] (155) at (24.75, -7) {};
		\node [style=none] (156) at (27.25, -7) {};
	\end{pgfonlayer}
	\begin{pgfonlayer}{edgelayer}
		\draw [bend left] (0.center) to (4);
		\draw [bend left] (4) to (6.center);
		\draw [bend left] (4) to (1.center);
		\draw [bend left, looseness=1.25] (12.center) to (4);
		\draw [bend left, looseness=1.25] (4) to (13.center);
		\draw [bend right=15] (4) to (14.center);
		\draw [bend right] (15.center) to (17);
		\draw [bend right] (17) to (18.center);
		\draw [bend left, looseness=1.25] (17) to (16.center);
		\draw [bend left] (21.center) to (17);
		\draw [bend left] (17) to (22.center);
		\draw [bend right, looseness=1.25] (17) to (23.center);
		\draw [bend right] (4) to (17);
		\draw [bend left] (4) to (17);
		\draw [in=90, out=0] (25.center) to (27);
		\draw [in=-180, out=90] (27) to (28.center);
		\draw [in=90, out=0] (31.center) to (27);
		\draw [in=-180, out=90] (27) to (32.center);
		\draw [in=0, out=-90] (27) to (42.center);
		\draw [in=270, out=0] (40.center) to (27);
		\draw [in=-180, out=-90] (27) to (43.center);
		\draw [in=270, out=180] (41.center) to (27);
		\draw [bend right=45] (46) to (52.center);
		\draw [bend left=45] (46) to (47.center);
		\draw [in=180, out=45] (46) to (51.center);
		\draw (46) to (53);
		\draw (53) to (50.center);
		\draw (52.center) to (56.center);
		\draw (51.center) to (55.center);
		\draw (47.center) to (54.center);
		\draw [bend right=45] (58) to (63);
		\draw [bend left=45] (58) to (59);
		\draw [in=180, out=45] (58) to (62);
		\draw (58) to (64.center);
		\draw (64.center) to (61.center);
		\draw (63) to (67.center);
		\draw (62) to (66.center);
		\draw (59) to (65.center);
		\draw [bend right=45] (79) to (84);
		\draw [bend left=45] (79) to (80);
		\draw [in=180, out=60] (79) to (83);
		\draw (79) to (85.center);
		\draw (85.center) to (82.center);
		\draw (84) to (88.center);
		\draw (83) to (87.center);
		\draw (80) to (86.center);
		\draw [bend right=45] (90) to (95.center);
		\draw [bend left=45] (90) to (91.center);
		\draw [in=180, out=60] (90) to (94.center);
		\draw (90) to (96);
		\draw (96) to (93.center);
		\draw (95.center) to (99.center);
		\draw (94.center) to (98.center);
		\draw (91.center) to (97.center);
		\draw (106) to (110.center);
		\draw (105) to (109.center);
		\draw (102) to (108.center);
		\draw [bend right=45] (112) to (117.center);
		\draw [bend left=45] (112) to (113.center);
		\draw [in=180, out=60] (112) to (116.center);
		\draw (112) to (118);
		\draw (117.center) to (121.center);
		\draw (116.center) to (120.center);
		\draw (113.center) to (119.center);
		\draw [bend left] (122.center) to (124);
		\draw [bend left] (124) to (123.center);
		\draw (124) to (125);
		\draw [bend left] (125) to (126.center);
		\draw [bend right] (125) to (127.center);
		\draw (131.center) to (129);
		\draw (129) to (130);
		\draw (130) to (132.center);
		\draw (135.center) to (133);
		\draw (133) to (134);
		\draw (134) to (136.center);
		\draw (129) to (134);
		\draw (130) to (133);
		\draw (138.center) to (140);
		\draw (140) to (139.center);
		\draw (144.center) to (150);
		\draw (150) to (151);
		\draw (151) to (145.center);
		\draw (155.center) to (156.center);
		\draw (152.center) to (153.center);
	\end{pgfonlayer}
\end{tikzpicture}
\centering
\caption{A set of the basic rewriting rules of ZX-calculus \cite{kissinger21}, where Greek letters denote arbitrary real variables, $[0,2\pi)$, and Latin letters denote arbitrary boolean variables, $\{0,1\}$. Note that the rules still apply if all the spider colours are inverted. These rules are known, in column-major order, as (a) spider fusion, (b) colour change, (c) $\pi$-commutation, (d) bialgebra, (e) state copy (where $n$ is the number of output edges), (f) identity removal, (g) Hadamard cancellation.}
\label{fig:basicrules}
\end{figure}
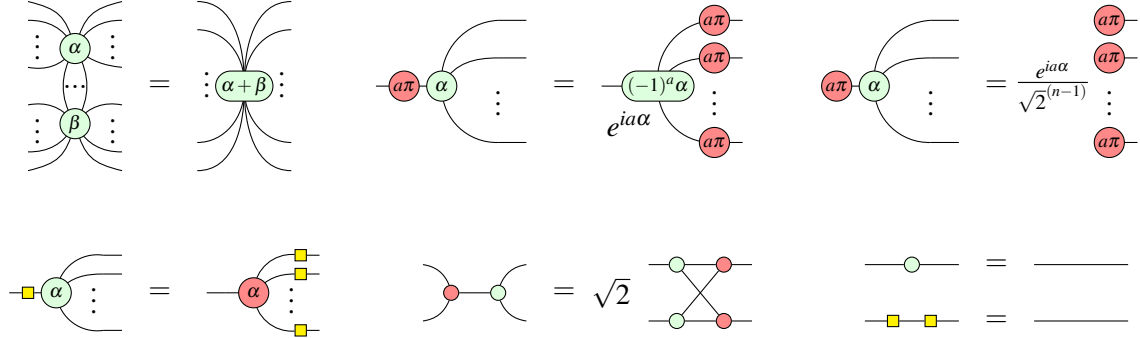

The major benefit of ZX-calculus is that it contains a set of well-defined \textit{rewriting rules} which outline how certain structures within ZX-diagrams may be equivalently written in simpler forms - allowing a circuit to be simplified while maintaining its behaviour. In particular, the fundamental rewrite rules are as outlined in figure \ref{fig:basicrules}.

\begin{figure}[h]
\begin{tikzpicture}
	\begin{pgfonlayer}{nodelayer}
		\node [style=Z phase dot] (4) at (1.5, 0) {$\;\pm\frac{\pi}{2}\;$};
		\node [style=none] (8) at (3, 0.25) {$\vdots$};
		\node [style=none] (10) at (6.5, 0) {$=$};
		\node [style=Z phase dot] (11) at (0, -1) {$\alpha_{2}$};
		\node [style=Z phase dot] (12) at (3, -1) {$\alpha_{3}$};
		\node [style=Z phase dot] (13) at (0, 1) {$\alpha_{1}$};
		\node [style=Z phase dot] (14) at (3, 1) {$\alpha_{n}$};
		\node [style=none] (15) at (2.5, 2) {};
		\node [style=none] (16) at (3.5, 2) {};
		\node [style=none] (17) at (3, 2) {$\cdots$};
		\node [style=none] (18) at (-0.5, 2) {};
		\node [style=none] (19) at (0.5, 2) {};
		\node [style=none] (20) at (0, 2) {$\cdots$};
		\node [style=none] (21) at (2.5, -2) {};
		\node [style=none] (22) at (3.5, -2) {};
		\node [style=none] (23) at (3, -2) {$\cdots$};
		\node [style=none] (24) at (-0.5, -2) {};
		\node [style=none] (25) at (0.5, -2) {};
		\node [style=none] (26) at (0, -2) {$\cdots$};
		\node [style=none] (28) at (17.25, 0.25) {$\vdots$};
		\node [style=Z phase dot] (29) at (14.5, -1) {$\;\alpha_{2}\mp\frac{\pi}{2}\;$};
		\node [style=Z phase dot] (30) at (17.5, -1) {$\;\alpha_{3}\mp\frac{\pi}{2}\;$};
		\node [style=Z phase dot] (31) at (14.5, 1) {$\;\alpha_{1}\mp\frac{\pi}{2}\;$};
		\node [style=Z phase dot] (32) at (17.5, 1) {$\;\alpha_{n}\mp\frac{\pi}{2}\;$};
		\node [style=none] (33) at (17, 2) {};
		\node [style=none] (34) at (18, 2) {};
		\node [style=none] (35) at (17.5, 2) {$\cdots$};
		\node [style=none] (36) at (14, 2) {};
		\node [style=none] (37) at (15, 2) {};
		\node [style=none] (38) at (14.5, 2) {$\cdots$};
		\node [style=none] (39) at (17, -2) {};
		\node [style=none] (40) at (18, -2) {};
		\node [style=none] (41) at (17.5, -2) {$\cdots$};
		\node [style=none] (42) at (14, -2) {};
		\node [style=none] (43) at (15, -2) {};
		\node [style=none] (44) at (14.5, -2) {$\cdots$};
		\node [style=none] (45) at (10.5, 0) {$e^{\pm{i\frac{\pi}{4}}} \sqrt{2}^{\frac{(n-1)(n-2)}{2}}$};
		\node [style=Z phase dot] (46) at (1.25, -6) {$k\pi$};
		\node [style=none] (47) at (2.5, -5.75) {$\vdots$};
		\node [style=none] (48) at (5.5, -6.5) {$=$};
		\node [style=Z phase dot] (51) at (-2.75, -5) {$\alpha_{1}$};
		\node [style=none] (56) at (-3.25, -4) {};
		\node [style=none] (57) at (-2.25, -4) {};
		\node [style=none] (58) at (-2.75, -4) {$\cdots$};
		\node [style=Z phase dot] (65) at (-1.25, -6) {$j\pi$};
		\node [style=Z phase dot] (66) at (-2.75, -7) {$\alpha_{n}$};
		\node [style=none] (67) at (-3.25, -8) {};
		\node [style=none] (68) at (-2.25, -8) {};
		\node [style=none] (69) at (-2.75, -8) {$\cdots$};
		\node [style=Z phase dot] (70) at (2.5, -5) {$\gamma_{1}$};
		\node [style=none] (71) at (2, -4) {};
		\node [style=none] (72) at (3, -4) {};
		\node [style=none] (73) at (2.5, -4) {$\cdots$};
		\node [style=Z phase dot] (74) at (2.5, -7) {$\gamma_{l}$};
		\node [style=none] (75) at (2, -8) {};
		\node [style=none] (76) at (3, -8) {};
		\node [style=none] (77) at (2.5, -8) {$\cdots$};
		\node [style=Z phase dot] (78) at (0, -7.25) {$\beta_{1}$};
		\node [style=none] (79) at (-0.75, -8) {};
		\node [style=none] (80) at (0.75, -8) {};
		\node [style=none] (81) at (0, -8) {$\cdots$};
		\node [style=Z phase dot] (82) at (0, -9.5) {$\beta_{m}$};
		\node [style=none] (83) at (-0.75, -10.25) {};
		\node [style=none] (84) at (0.75, -10.25) {};
		\node [style=none] (85) at (0, -10.25) {$\cdots$};
		\node [style=none] (86) at (-2.75, -5.75) {$\vdots$};
		\node [style=none] (88) at (0, -8.25) {$\vdots$};
		\node [style=none] (89) at (9, -6.5) {$(-1)^{jk}\sqrt{2}^{E}$};
		\node [style=none] (91) at (20.5, -5.75) {$\vdots$};
		\node [style=Z phase dot] (92) at (12.5, -5) {$\;\alpha_{1}+k\pi\;$};
		\node [style=none] (93) at (12, -4) {};
		\node [style=none] (94) at (13, -4) {};
		\node [style=none] (95) at (12.5, -4) {$\cdots$};
		\node [style=Z phase dot] (97) at (12.5, -7) {$\;\alpha_{n}+k\pi\;$};
		\node [style=none] (98) at (12, -8) {};
		\node [style=none] (99) at (13, -8) {};
		\node [style=none] (100) at (12.5, -8) {$\cdots$};
		\node [style=Z phase dot] (101) at (20.5, -5) {$\;\gamma_{1}+j\pi\;$};
		\node [style=none] (102) at (20, -4) {};
		\node [style=none] (103) at (21, -4) {};
		\node [style=none] (104) at (20.5, -4) {$\cdots$};
		\node [style=Z phase dot] (105) at (20.5, -7) {$\;\gamma_{l}+j\pi\;$};
		\node [style=none] (106) at (20, -8) {};
		\node [style=none] (107) at (21, -8) {};
		\node [style=none] (108) at (20.5, -8) {$\cdots$};
		\node [style=Z phase dot] (109) at (16.5, -8) {$\;\beta_{1}+(j+k+1)\pi\;$};
		\node [style=none] (110) at (15.75, -8.75) {};
		\node [style=none] (111) at (17.25, -8.75) {};
		\node [style=none] (112) at (16.5, -8.75) {$\cdots$};
		\node [style=Z phase dot] (113) at (16.5, -10.25) {$\;\beta_{m}+(j+k+1)\pi\;$};
		\node [style=none] (114) at (15.75, -11) {};
		\node [style=none] (115) at (17.25, -11) {};
		\node [style=none] (116) at (16.5, -11) {$\cdots$};
		\node [style=none] (117) at (12.5, -5.75) {$\vdots$};
		\node [style=none] (118) at (16.5, -9) {$\vdots$};
	\end{pgfonlayer}
	\begin{pgfonlayer}{edgelayer}
		\draw [style=hadamard edge] (4) to (13);
		\draw [style=hadamard edge] (4) to (14);
		\draw [style=hadamard edge] (4) to (12);
		\draw [style=hadamard edge] (4) to (11);
		\draw (15.center) to (14);
		\draw (16.center) to (14);
		\draw (13) to (18.center);
		\draw (13) to (19.center);
		\draw (11) to (24.center);
		\draw (11) to (25.center);
		\draw (12) to (21.center);
		\draw (12) to (22.center);
		\draw (33.center) to (32);
		\draw (34.center) to (32);
		\draw (31) to (36.center);
		\draw (31) to (37.center);
		\draw (29) to (42.center);
		\draw (29) to (43.center);
		\draw (30) to (39.center);
		\draw (30) to (40.center);
		\draw [style=hadamard edge] (31) to (29);
		\draw [style=hadamard edge] (29) to (30);
		\draw [style=hadamard edge] (30) to (32);
		\draw [style=hadamard edge] (32) to (31);
		\draw [style=hadamard edge] (31) to (30);
		\draw [style=hadamard edge] (32) to (29);
		\draw (51) to (56.center);
		\draw (51) to (57.center);
		\draw (66) to (67.center);
		\draw (66) to (68.center);
		\draw (70) to (71.center);
		\draw (70) to (72.center);
		\draw (74) to (75.center);
		\draw (74) to (76.center);
		\draw (78) to (79.center);
		\draw (78) to (80.center);
		\draw (82) to (83.center);
		\draw (82) to (84.center);
		\draw [style=hadamard edge] (65) to (51);
		\draw [style=hadamard edge] (65) to (66);
		\draw [style=hadamard edge] (65) to (46);
		\draw [style=hadamard edge] (46) to (70);
		\draw [style=hadamard edge] (46) to (74);
		\draw [style=hadamard edge, bend right] (65) to (78);
		\draw [style=hadamard edge, bend right] (78) to (46);
		\draw [style=hadamard edge, in=165, out=-105, looseness=0.75] (65) to (82);
		\draw [style=hadamard edge, in=285, out=15, looseness=0.75] (82) to (46);
		\draw (92) to (93.center);
		\draw (92) to (94.center);
		\draw (97) to (98.center);
		\draw (97) to (99.center);
		\draw (101) to (102.center);
		\draw (101) to (103.center);
		\draw (105) to (106.center);
		\draw (105) to (107.center);
		\draw (109) to (110.center);
		\draw (109) to (111.center);
		\draw (113) to (114.center);
		\draw (113) to (115.center);
		\draw [style=hadamard edge] (92) to (101);
		\draw [style=hadamard edge] (92) to (105);
		\draw [style=hadamard edge] (101) to (97);
		\draw [style=hadamard edge] (92) to (109);
		\draw [style=hadamard edge] (92) to (113);
		\draw [style=hadamard edge] (101) to (109);
		\draw [style=hadamard edge] (101) to (113);
		\draw [style=hadamard edge] (97) to (109);
		\draw [style=hadamard edge] (97) to (113);
		\draw [style=hadamard edge] (105) to (109);
		\draw [style=hadamard edge] (105) to (113);
	\end{pgfonlayer}
\end{tikzpicture}
\centering
\caption{Two important derived ZX-calculus rewrite rules \cite{kissinger21}, namely (top) local complementation and (bottom) pivoting. Here, $E=(n-1)m+(l-1)m+(n-1)(l-1)$.}
\label{fig:derivedrules}
\end{figure}
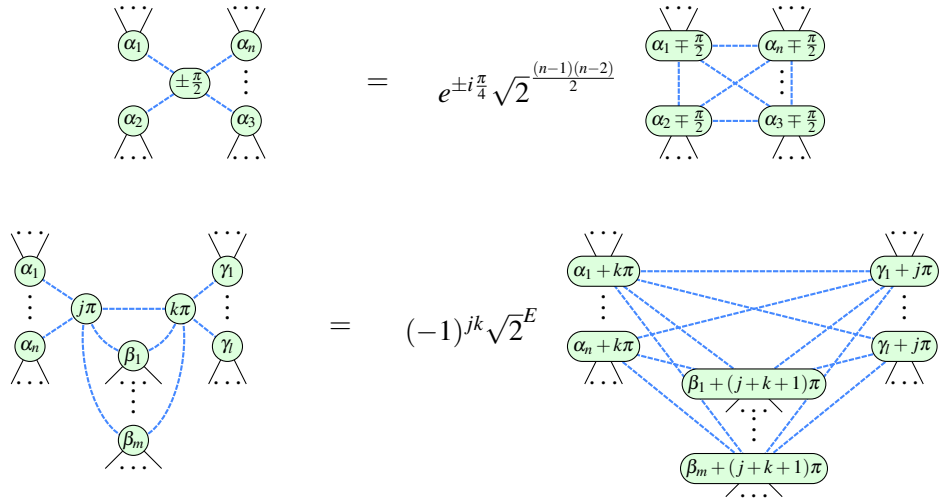

From these basic rules, one may derive a number of more complex rules. In particular, \textit{local complementation} and \textit{pivoting}, outlined in figure \ref{fig:derivedrules}, prove to be extremely useful in reducing ZX-diagrams. A more thorough collection of derived rules may be found in \cite{wetering20}.

A particular subset of ZX-diagrams are those with no open input or output wires. These are known as \textit{scalar} diagrams, and the rules above are sufficient to reduce such diagrams, of the Clifford gateset, to complex scalars, making use of the following scalar relations:

\ctikzfig{scalars}

\subsection{Classical simulation}
\label{subsec:classicsim}

To verify the behaviour of complex quantum algorithms, especially when present-day quantum hardware is insufficient to run them, one may turn to classical simulation. There are two classes of this - namely \textit{strong} simulation and \textit{weak} simulation. For the purposes in this paper, we will focus on the former, wherein the aim is to determine the probability of a specific measurement outcome. In short, this may be done with ZX-calculus by `plugging' the inputs and outputs with spiders corresponding to the desired states and measurements to produce a \textit{scalar} ZX-diagram, which may then be fully reduced to a simple complex scalar value. This result will represent its amplitude, $A$, and thus relates to its probability, $|A|^2$.

Notably, Cliford diagrams (those restricted to phases of $\frac{n\pi}{2}$) are very efficient to simulate classically. This manifests itself in ZX-calculus in that such diagrams, given no open inputs or outputs, may be fully reduced to a scalar via the rewriting rules highlighted above. Clifford+T diagrams, on the other hand, are notoriously inefficient to simulate classically. While the rewriting rules may remove a number of T-gates from a ZX-diagram, it is generally unable to remove them all, meaning such diagrams are unable to fully reduce to a scalar. To deduce the scalar amplitude in such cases then, one must make use of decompositions to translate the Clifford+T diagram into a sum of efficiently-reducible Clifford diagrams. The inefficiency lies in the fact that the number of such summand Clifford terms one attains in place of a Clifford+T diagram scales exponentially with the number of T-gates. This is typically quantified with a parameter $\alpha$ which represents the efficiency of the precise decomposition(s) used, given it translates a Clifford+T diagram of $t$ T-gates into a sum of $2^{\alpha t}$ Clifford terms. So, smaller values of $\alpha$ describe more efficient decompositions.

\begin{figure}[h]
\input{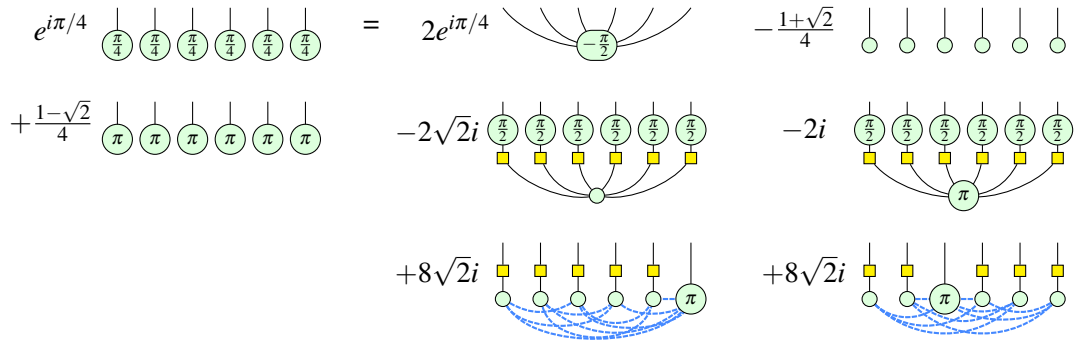}
\centering
\caption{The BSS decomposition \cite{BSS}, relating a set of 6 T-gates to a sum of 7 Clifford terms, expressed in ZX-diagrams as per \cite{kissinger21}.}
\label{fig:bssdecomp}
\end{figure}

For instance, the decomposition presented by Bravyi, Smith, and Smolin \cite{BSS}, and expressed in ZX-calculus terms by Kissinger and van de Wetering \cite{kissinger21}, allows sets of 6 T-gates to be replaced with a sum of 7 Clifford terms, according to figure \ref{fig:bssdecomp}. As such, this ``BSS'' decomposition scales as $7^{t/6}\approx2^{0.468t}$, hence $\alpha\approx0.468$. The current state of the art T-decompositions, meanwhile, achieve $\alpha\approx0.396$ \cite{Qassim2021improvedupperbounds}.

As outlined in \cite{kissinger21}, after every decomposition, some further reduction in T-count may be facilitated by Clifford simplification via the rewriting rules, such that the $2^{\alpha t}$ actually represents an upper-bound estimate of the number of terms ultimately produced. Moreover, in their work they essentially select groups of 6 T-gates for decomposition arbitrarily, without regard for whether certain groupings may be more likely to allow further simplification after decomposition. Lastly, while their method relies primarily on the BSS decomposition, when the number of T-gates remaining in any graph falls below 6, they fall back on other known decompositions.

\section{Method}
\label{sec:method}

\subsection{Graph cutting}
\label{subsec:graphcut}

While Clifford+T diagrams are typically reduced via such T-gate decompositions as that of Bravyi, Smith, and Smolin, highlighted above, paying close attention to the structures inherent in a given ZX-diagram can reveal that blindly decomposing T-gates in an essentially random order is not necessarily optimal. In fact, decomposing conveniently positioned \textit{Clifford} spiders - even with an apparently less efficient decomposition - can actually produce more efficient results.

From the definitions of Z- and X- spiders in section \ref{subsec:zxcalc}, one can infer the basic relation:

\ctikzfig{spider_decomp_abs}

Note that the equality here is up to some global scalar factor (neglected for brevity). This acts as a very simple decomposition for any arbitrary spider. Herein, the act of applying this particular decomposition will be referred to as `\textit{cutting}' the graph, as structurally it behaves like physically slicing a vertex from its edges. Superficially, this may not seem particularly useful as, applied to T-gates, this doubles the number of terms for each one (hence has a very poor efficiency of $\alpha=1$). However, by cutting appropriate spiders (not necessarily even T-like spiders), this can allow some T-gates to cancel, along both paths of the decomposition.

Specifically, consider a subgraph consisting of two T-like spiders sandwiched between a CNOT. Cutting one end of the CNOT (the side opposite the T-gates) will produce two terms, like so:

\ctikzfig{picom_01}

An observant reader may notice that both of these terms, after a little simplification via the rewriting rules, may fuse their T-gates into a Clifford. The left-hand term, for instance, simplifies as follows:

\ctikzfig{picom_01a}

The right-hand term simplifies similarly, except utilising $\pi$-commutation in place of identity removal:

\ctikzfig{picom_01b}

Consequently, via a cut of a Clifford spider the T-count has been reduced by 2. This gives a decomposition efficiency of $\alpha=0.5$. Moreover, if the green end of the CNOT had itself taken a T-like phase, then that too would have been removed (specifically, converted into a scalar factor directly by the decomposition). In that case, one would have observed a better efficiency of $\alpha=1/3$.

\subsection{CNOT-grouping}
\label{subsec:cnotgrouping}

One may further recognise that it is possible for any number of these T-CNOT-T arrangements to be aligned such that their respective CNOTs may fuse. In such cases, each pair of T-like spiders may fuse to a Clifford as above, though still just requiring one vertex cut overall. As a simple example, the following shows how two sets might be grouped and reduced in this way:

\ctikzfig{picom_grouped}

Here, the T-count has been reduced by 4, still at the expense of just one vertex cut (hence 2 Clifford terms). This corresponds to $\alpha=0.25$. It should be clear from this point that this reasoning could be extended ad infinitum, with arbitrarily many T-CNOT-T subgraphs being fusible and hence arbitrarily many T-gates being reducible to Clifford at the expense of just one cut, giving $\alpha\rightarrow0$.

A similar concept to this was recognised in \cite{Codsi2022Masters} (albeit notated in a different fashion), wherein, for a single decomposition utilising $\pi$-commutations, T-count reductions of up to 286 were found on SAT counting \cite{berent22,debeaudrap21} ZX-diagrams (giving $\alpha\approx0.0035$). However, this work utilised a very simple heuristic amounting to prioritising decompositions of T-gates with the maximum number of immediately relevant connections. Perhaps surprisingly, however, this na\"{i}ve heuristic seldom produces the most efficient solutions and indeed can often be extremely suboptimal. Moreover, their approach to recognising instances where these cuts are applicable is essentially limited to those where the decomposition is to be applied to a T-like spider, rather than any arbitrarily-phased spider as per the means outlined above. And lastly, by acting on ZX-diagrams \textit{after} they have undergone full Clifford simplification, their approach risks losing much of the graph's structure and thus missing relevant patterns.

\subsection{Cutting in tiered structures}
\label{subsec:tiered}

Just as a CNOT may be the only thing standing in the way of two or more T-like spiders from fusing, so too may a CNOT be the only obstacle preventing some T-CNOT-T sandwiches from grouping. Consider, as a prime example, the structured circuit that follows, where the Z-spiders here have been labelled for easy reference:

\ctikzfig{tiered_01b}

Immediately, one might recognise 7 T-CNOT-T sandwiches, centred on vertices $\#2$, $\#4$, $\#6$, $\#9$, $\#11$, $\#13$, and $\#15$, respectively. Notably, however, there is some clashing here, in that - for example - the sandwich centred on $\#2$ and that centred on $\#4$ are mutually exclusive, as they a share one of their T-gates. Indeed, there are a number of possible ways of reducing this circuit via the means outlined so far in this paper.

For example, cutting vertices $\#4$, $\#9$, and $\#13$ would allow 6 T-gates to reduce to Clifford. Alternatively, cutting $\#2$, $\#6$, $\#11$, and $\#15$ would reduce all 8 T-gates, at the cost of 4 cuts (hence $\alpha=0.5$). But, the best solution here would be to firstly cut vertex $\#8$ - even though this doesn't immediately allow any T-gates to reduce - as this then allows vertices $\#2$, $\#6$, $\#11$, and $\#15$ to fuse into one. Cutting this newly fused vertex then would allow all 8 T-gates to reduce. So, this solution would reduce the T-count entirely (all 8 T-gates) at the cost of just 2 cuts (hence $\alpha=0.25$). And of course, this reasoning could be extended for higher tiers, where the optimal initial cuts are two, three, or more, steps away from any T-gates.

Similarly, there may be instances where \textit{multiple} CNOTs are directly blocking some set of T-gates from fusing. In such cases, the likelihood that the cutting all of those CNOTs would be worthwhile to reduce the T-gates they block will be determined by the ratio of the CNOTs to T-gates involved, as well as whether some of those CNOTs are considered worthwhile cuts in their own right, with regard to any other T-gates that they alone may be blocking. We can call such groups of spiders `spousal', with respect to the children spiders they are collectively blocking from reducing.

Evidently, therefore, selecting which vertices to cut, and indeed in which order, is a very intricate task (though the former is more important as suboptimality in the latter can be corrected for via a slight modification and parametric analysis, as detailed in appendix \ref{app:para-analysis}). Na\"{i}vely tackling this problem via an exhaustive, brute force, approach would require checking the reduction achieved by every possible combination of vertex cuts. This is obviously infeasible for large-scale graphs, as the time complexity scales exponentially with the number of vertices. Consequently, a heuristic approach is desired. On this note, as we have shown, simply prioritising vertices which are directly blocking the most number of T-like pairs from fusing is not generally optimal. Rather, it is preferential to look at the whole picture and determine the optimal cuts on higher tiers.

\subsection{Optimised cutting procedure}
\label{subsec:proc}

The solution we propose is a procedure based on assigning weights to vertices, determined by how many T-like gates they are preventing from fusing to Clifford, and then propagating these weights through any neighbours which are then preventing weighted vertices from fusing, and so on up the tiers. Particular care is given to balance the weightings appropriately, especially in places where multiple `spousal' cuts are required to facilitate a fusion of their children.

Note that we may label the weight of a vertex (i.e. spider) $v$, for a particular tier $t$, as $w_v^t$, such that, for instance, $w_{12}^2$ refers to the weight of vertex $\#12$ with respect to tier $\#2$. Given this, the procedure steps are as follows:
\begin{enumerate}
  \item Partially simplify the circuit such that any instances of spider fusion are applied (so no like-coloured spiders remain directly connected via a solid edge) and any $\pi$-phase spiders are pushed to one side or into CNOTs via the $\pi$-commutation and fusion rules. Then, let $t=0$ and assign an initial weight of $0$ to every spider (i.e. let $w_v^0=0$ $\forall v$).
  \item \label{proc-step-T} For any pair of T-like spiders that are separated by $k$ (that is, one or more) CNOTs, add $2/k$ to the weights of the opposite ends of each of those CNOTs. (For any given CNOT, take care not to count a particular T-spider more than once.) Now, for instance, any CNOT that is preventing $2$ T-spiders from fusing to Clifford will have a corresponding weight of $2$, and any CNOT preventing, collectively, $4$ T-spiders from reducing to Cliffords will have a corresponding weighting of $4$, etc. Similarly, if, for instance, $3$ CNOTs are collectively blocking a single pair of T-spiders from fusing, then each will have a weighting of $2/3$.
  \item \label{proc-step-iterate} Increment $t\leftarrow t+1$. Then, similar to step \ref{proc-step-T}, for any \textit{weighted} vertex $v$ of the previous tier (i.e. any $v$ for which $w_v^{t-1}\geq0$) that is separated from fusing with another weighted vertex of \textit{any} lower tier (i.e. any $v$ for which $w_v^{u}\geq0$ for any $u<t$) by $k$ (that is, one or more) CNOTs, add $\gamma(w_v^{t-1})/k$ to the weight of the opposite ends of each of those CNOTs. Here, the $\gamma$ function normalises a given weighting to the range $[0,1]$, such that (crudely speaking) $0$ roughly means ``very unlikely to be a worthwhile cut'' and $1$ ``very likely to be a worthwhile cut'': $\gamma(w):=\min{(\frac{w}{2},1)}$.
  \item Repeat step \ref{proc-step-iterate} until no new changes are made (that is, until no weightings, $w_v^t$, are found for any $v$, given the current $t$). At this point, one will be left with weightings for every vertex, for every tier ($w_v^t$ $\forall v,t$). From this, one can trivially extract the relevant data, namely, for every vertex $v$, its \textit{maximum} weight $w_v^t$ (for any $t$) and, respectively, the maximum $t$ for which the vertex has a weight (i.e. largest $t$ for which $w_v^t\geq0$). One may label the maximum weight of a vertex, $W_v$, and its maximum tier for which it has a non-zero weight, $T_v$.
  \item \label{proc-step-cutbest} Among the vertices with the largest recorded $T$, namely $T_{max}$, select the one with the greatest max weight $W_v$ (i.e. select vertex $V$ s.t. $W_V\geq W_v$ $\forall v$ s.t. $T_v=T_V=T_{max}$). (Note that for this step only, we may temporarily add $1$ to the weight of any vertex of a T-like phase.) If this weight is below $2$ (hence $\gamma(W_V)\leq1$, implying the cut would not likely be worthwhile), then search instead among the vertices of the lower tier (i.e. for which $T_v=T_{max-1}$), until an appropriate vertex is found for which $W_V\geq2$. This is the vertex which the heuristic has concluded is likely an optimal choice to cut. As such, cut this vertex (i.e. decompose it into two branches as per section \ref{subsec:graphcut}).
  \item \label{proc-step-simpbranches} Partially simplify each branch, without compromising the graph structure. Specifically, push any new $\pi$-phase spiders to one side, and/or into CNOTs, via repeated applications of the $\pi$-commutation rule (and fusion), and thereafter apply the fusion rule until no like-coloured spiders remain connected via a solid edge. Moreover, when fusing weighted spiders, update their combined weight accordingly as the sum of their respective max weights (i.e. in fusing vertex $B$ into vertex $A$, the former is removed along with its recorded weightings and tier data and the latter is updated as $W_A\leftarrow W_A+W_B$). Similarly, the max tier of the newly fused vertex takes that of the larger of the two fused vertices (i.e. $T_A\leftarrow\max{(T_A,T_B)}$). Moreover, recalculate the weightings on any vertices whose children have been altered by this partial simplification. 
  \item Repeat steps \ref{proc-step-cutbest} and \ref{proc-step-simpbranches} until no further cuts are made (or the number of T-spiders $\leq2$). If any T-like spiders remain, then these may be decomposed via a typical T-decomposition (e.g. the BSS decomposition outlined in section \ref{subsec:classicsim}).
\end{enumerate}

Applying this procedure to a Clifford+T ZX-diagram will result in a set of Clifford ZX-diagrams whose sum equals the original. For scalar diagrams (those with no open input or output wires), each summand will simply be a scalar term, and thus the overall sum will be likewise \cite{kissinger21}.

Python code, based on the PyZX package \cite{github:pyzx,kissinger2020Pyzx}, that implements this procedure may be found at \url{https://github.com/mjsutcliffe99/ProcOptCut} \cite{github:procOptCut}, and a step-by-step illustrative example of this procedure in action is shown therein.

\section{Results}

\subsection{Complexity and efficiency}
\label{subsec:complexity}

The exact runtime complexity of the procedure is difficult to discern as it depends on the density of the graph (i.e. typical number of neighbours the vertices have) and its number of `tiers', $T_{max}$, (which is heuristically determined). Very reasonably, one could assume $T_{max}\ll V$ (for a non-trivial number of vertices, $V$), and so a crude upper-bound (taking, very unrealistically, maximal density) may be given by $O(V^2)$. Even this gross over-estimate of the runtime complexity shows that, compared to the number of stabiliser terms produced ($2^{\alpha t}$ for $t$ T-gates and some $\alpha<1$), the procedure's runtime is negligible, so that any improvements it offers to the number of the stabiliser terms can be taken as is.

To test the effectiveness of the proposed method, we explored how often - for circuits small enough to verify - it was able to find the \textit{most} optimal set of vertex cuts, rather than simply \textit{an} optimal set. Specifically, we considered many random, non-trivially structured (see appendix \ref{app:gen-rand-circs}) ZX-diagrams of $16$ internal Z-spiders or fewer and measured the number of stabiliser terms required to remove all T-spiders via the method outlined in section \ref{subsec:proc}. In each case, we also tested every possible combination of vertex cuts on the internal Z-spiders (applying the reasoning of appendix \ref{app:para-analysis} to correct for any suboptimal cut ordering). Thus, we could determine with certainty the most optimal set of vertex cuts (on Z-spiders) and correspondingly the number of stabiliser terms produced (or, by extension, the effective $\alpha$) and compare this, in each case, to the result achieved by our method. In this way, we observed that our method found the most optimal set of cuts possible (on Z-spiders) $71\%$ of the time. Even so, in every case in which the method failed to find the most optimal solution, it still invariably found a very good solution with an $\alpha$ usually only marginally above that of the best. It is worth noting, however, that we cannot be certain that this would translate to a similar success rate for much larger circuits (i.e. those too big to verify by brute force), where greater levels of structure (and indeed chaos) are possible. As such, to extend our analysis we compared our method also to that of the more typical approach of T-decompositions a la \cite{kissinger21}, of which we outline our results ahead.

\subsection{Experimental measurements for random circuits}

Applying the procedure outlined in section \ref{subsec:proc} to a Clifford+T ZX-diagram will produce an exact result (e.g. for computing the probability amplitude of a Clifford+T diagram for classical simulation). However, in benchmarking the efficiency of the method, one is more interested in the \textit{number} of stabiliser terms the original diagram is decomposed to, rather than the precise value of the sum of these terms. With this in mind, one can run a `\textit{blind}' version of the procedure to compute an (upper-bound) estimate of the number of resulting terms in linear time. As well as not providing the final numerical result, this blind version of the procedure also misses out on inter-step ZX-calculus simplifications, and hence typically overestimates the number of resulting terms (and thus is an upper-bound estimate). The appropriate modifications for this version of the procedure are outlined and justified in appendix \ref{app:estimate}.

Similarly, one can likewise calculate an upper-bound estimate of the number of terms produced after fully decomposing a Clifford+T diagram via the BSS-based approach of \cite{kissinger21}. This is much more simple to do as the BSS decomposition efficiency is known in advance, namely $\alpha\approx0.468$. Thus, an estimate of the number of resulting terms this method achieves for a given diagram of T-count $t$ is given by $2^{0.468t}$, which can be computed trivially in constant time. As before, this does not account for the inter-step ZX-calculus simplifications and hence is an approximate upper-bound estimate.

\begin{figure}[h]
  \begin{subfigure}{.5\textwidth}
  \centering
    \includegraphics[width=1\linewidth]{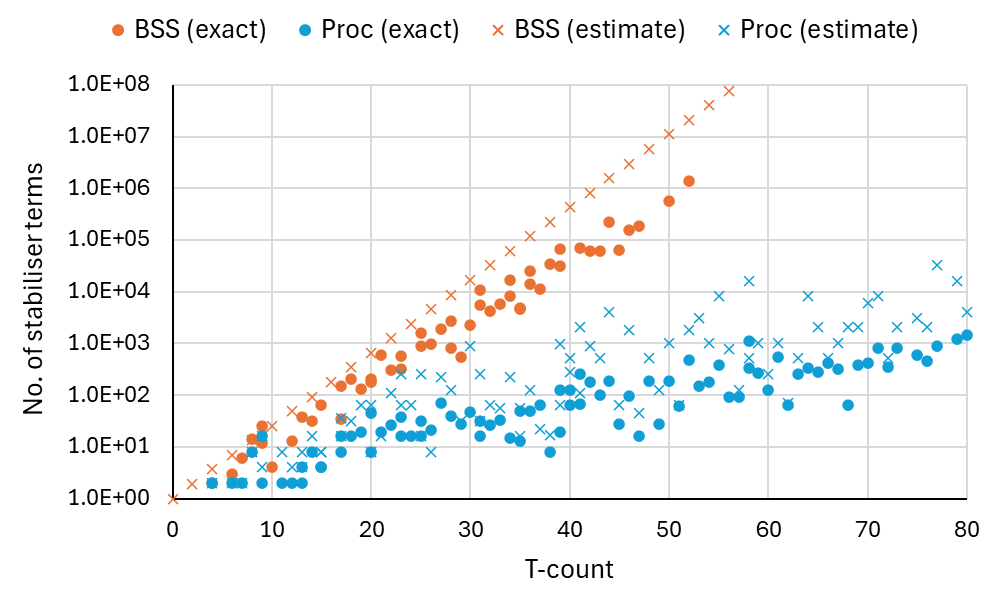}
    \caption{Number of stabiliser terms versus T-count}
    \label{fig:res-terms}
  \end{subfigure}%
  \begin{subfigure}{.5\textwidth}
  \centering
    \includegraphics[width=1\linewidth]{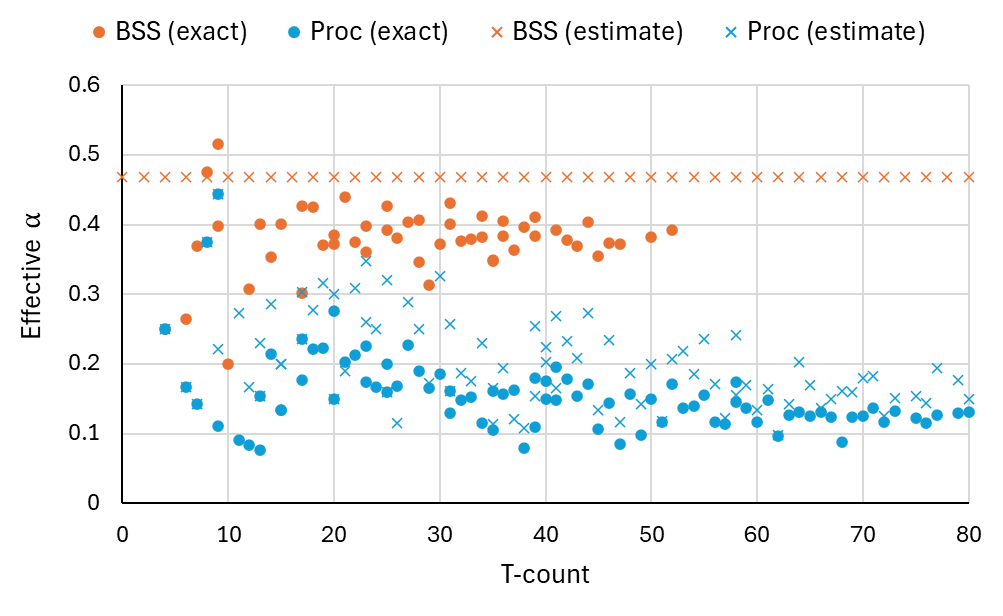}
    \caption{Effective efficiency $\alpha$ versus T-count}
    \label{fig:res-alpha}
  \end{subfigure}
  \caption{(a) The number of stabiliser terms, $n$, produced after decomposing all $t$ T-gates, and correspondingly (b) the effective overall decomposition efficiency $\alpha$ (given by $\alpha=\frac{1}{t}\log_2{n}$) for numerous randomly generated (see appendix \ref{app:gen-rand-circs}) pseudo-structured Clifford+T circuits of various T-counts. In each case, we measure experimentally the exact results, as well as approximate upper-bound estimations, achieved by both the conventional (``BSS'') method of \cite{kissinger21} and the procedural approach (``proc'') presented in this paper.}
  \label{fig:results}
\end{figure}

For various randomly generated ZX-diagrams (constructed from CNOTs, $\frac{n\pi}{4}$ phase gates, and Toffolis as per appendix \ref{app:gen-rand-circs}), we measured the number of stabiliser terms required to remove all T-spiders via the approach of Kissinger and van de Wetering \cite{kissinger21} versus the method we propose in section \ref{subsec:proc}. For computationally feasible T-counts (those computable within a few minutes), we can compute these exactly with both methods, and for larger (computationally infeasible) T-counts, we instead rely on the upper-bound estimations outlined above. For each of these measurements, we can also infer the overall effective decomposition efficiency, $\alpha$, by simply taking $\alpha=\frac{1}{t}\log_2{n}$, where $t$ is the initial T-count of the circuit (after full Clifford simplification) and $n$ is the number of stabiliser terms to which it is reduced. The results are illustrated in figure \ref{fig:results}. (Note that the initial T-counts are all taken as that achieved after full Clifford simplification.)

We observe that our procedural method, applied to circuits of this type, is very effective at minimising the number of resulting stabiliser terms, compared to relying predominantly on na\"{i}ve applications of the BSS decomposition. The exact effectiveness of our method can vary quite substantially from circuit to circuit, depending on how much structure they embed. The direct BSS approach, meanwhile, tends to be a lot more consistent, though can also vary somewhat depending on how much inter-step simplification it is able to undertake. Regardless, the procedural method very consistently shows magnitudes of improvement on these circuits, allowing above even double the T-counts to be computed within the same time frame.

This is reflected too in the effective $\alpha$ measurements, where, for circuits of this type, our method offers typically $0.1\lesssim\alpha\lesssim0.2$ for such pseudo-structured circuits. The results are less remarkable on trivially small T-counts as such small circuits have little room for much structure of which to take advantage. Evidently, therefore, this method is, as intended, very well optimised for decomposing Clifford+T circuits that exhibit some structure, enabling such circuits of much larger T-counts to be computed within reasonable time frames.

For full context, it is worth remembering two points when considering these results. Firstly, the random circuits generated for these experiments were those with some inherent localised structural elements. And secondly, while the runtime of the relevant steps of the procedure, after each decomposition, is very quick, it does nevertheless contribute time to each term that is not present in the BSS method, so that the difference in numbers of terms produced by the two methods does not translate 1-to-1 to a difference in runtime. In other words, computing $n$ terms via the procedural method may be a little slower than computing $n$ terms via the BSS method (although this difference will not be more than some small factor, so that the improvement offered by the reduced number of terms still vastly outweighs this offset).

All of the above experiments may be repeated from the corresponding Jupyter notebook, hosted on Github \cite{github:procOptCut}.

\section{Conclusion}

We demonstrated how a procedure could be designed to analyse the structure of any given Clifford+T quantum circuit in order to determine an optimised set of vertex cuts to efficiently decompose it to a sum of (classically computable) Clifford terms. This is contrary to the more conventional approach \cite{kissinger21}, which applies decompositions essentially arbitrarily, without such regard to the specific structures involved.

Specifically, we show that our method is very effective at finding optimal sets of vertex cuts, with a $71\%$ success rate at finding the \textit{most} optimal set when applied to random small semi-structured circuits. We further show that our method is very efficient at decomposing even larger such circuits, consistently outperforming the more conventional approach \cite{kissinger21} by orders of magnitude, with regard to the number of resulting terms (and by extension the runtime). What this means in practise is that our method could allow for classical simulation, within a reasonable time frame, of Clifford+T circuits of more than double the T-count as could be achieved with the conventional methods. This is hugely relevant to verifying the behaviour of quantum software and hardware for present-day NISQ (noisy intermediate-scale quantum) devices \cite{preskill18}.

While already very effective, there are many ways in which the method outlined in this paper could be improved - many stemming from the rigid scope with which it applies and propagates weights. After all, T-gate fusion facilitated by cutting a CNOT is just one way in which the T-count of a circuit may be reduced. It would be worth considering also the potential of cuts to remove T-gates by pushing them into the scalar factor (e.g. via the state copy rule) or to partition small segments from the graph. The method ought also to consider simplification and weight propagation laterally, rather than solely through the lens of `tiers'. Indeed, the `partial simplification' strategy we utilise (to simplify while maintaining the structure) does not take into account some vital steps in the normal `full reduce' \cite{github:pyzx} function (namely pivoting - i.e. on CNOTs - and local complementation). We suspect this is largely the reason we didn't observe an even higher success rate in verifying how often the method was able to find the best solutions on small circuits. Moreover, a more robust analysis could determine appropriate weightings \textit{without} the need for the circuit to be expressed in a very rigid graph-like form, such that further simplification between steps could be enabled and, for instance, Toffolis could be expressed as phase gadgets (so that we are not restricted due to the arbitrary choice of which way around to decompose each Toffoli's control qubits). And lastly, of course, one could consider cuts on X-spiders as well as just Z-spiders (this might be particularly relevant if there are many Hadamards involved, resulting in many X-spiders of T-like phase).

There are also a number of ways in which this concept, more broadly, could be improved, such as developing newer and better heuristics - perhaps even different heuristics for different types of circuit (e.g. dense circuits, or those with many Toffolis, etc.). Nevertheless, we demonstrate very clearly how analysing the circuit structure and applying decompositions discriminately can offer vastly more efficient results than simply decomposing the T-spiders directly with a decomposition that has a better \textit{immediate} efficiency.

\nocite{*}
\bibliographystyle{eptcs}
\bibliography{dynamic-cutting}


\appendix

\section{Cut order correction}
\label{app:para-analysis}
It was shown in section \ref{subsec:tiered} that determining the \textit{order} in which to cut the vertices is apparently at least as important as determining \textit{which} vertices to cut. In the example showcased there, it seemed necessary for the optimal solution to cut vertex $\#8$ first, such that vertices $\#2$, $\#6$, $\#11$, and $\#15$ could then be fused and cut as one. It would appear that recognising the same vertices to cut but applying a different cut ordering (specifically cutting vertex $\#8$ \textit{last}) would achieve the same ends at the cost of $5$ cuts rather than $2$ (and hence a much less efficient $32$ stabiliser terms rather than $4$). However, with slight alteration (using the modified PyZX package of \cite{github:paramzx}) to denote the cuts parametrically, followed by some simple parametric analysis, a suboptimal cut ordering can effectively be corrected to its more optimal arrangement at a negligible cost to the runtime.

Firstly, one can parameterise the cutting decomposition of section \ref{subsec:graphcut} like so:

\ctikzfig{spider_decomp_para}

Now consider again the circuit shown in section \ref{subsec:tiered}, and imagine cutting first vertices $\#2$, $\#6$, $\#11$, and $\#15$. Writing this in the parameterised form - needing only one parameterised graph (rather than $16$ evaluated graphs) - leads to the following, after only some trivial spider fusion:

\ctikzfig{tiered_01_paracut}

Cutting lastly vertex $\#8$ results in the following:

\ctikzfig{tiered_01_paracut2}

Here we have a graph in 5 parameters ($a,b,c,d,e$), containing 3 free nodes (legless spiders). As the parameters are boolean, it necessarily follows that each of these free nodes can be replaced with a scalar factor of $0$ or $2$, such as follows:

\ctikzfig{free_node_para}

Moreover, for any combination of parameter values that results in a scalar factor of $0$ in one or more of these free nodes, one can ignore the entire corresponding graph (as the whole graph then becomes $0$). In other words, one is only interested in the sets of parameter values which result in non-zero scalar factors for all free nodes. So, given the equation above, the leftmost free node must equal $2$ and thus $a\oplus b\oplus e = 0$. Rearranging this for, say, $a$ gives: $a=b\oplus e$. And now, every instance of $a$ throughout the parameterised graph can be substituted out for $b\oplus e$, thus reducing the number of parameters from $5$ to $4$. Repeating this reasoning for the remaining two free nodes finds that $b=c\oplus e$ and $c=d\oplus e$, resulting in all parameters being reduced to some combination of $d$ and $e$.

Consequently, while only needing to reason on one graph, the number of parameters has been reduced from the $5$ attained via a suboptimal cut ordering to the most optimal $2$. Indeed, this graph is now equivalent to what would have been attained if the more optimal cut ordering had been adopted (namely cutting vertex $\#8$ first and then fusing the remaining 4 before cutting them as one). Having effectively corrected suboptimal cut ordering, one can then expand the parameterised graph out into its (in this case $4$) distinct evaluated graphs and proceed with simplification as in section \ref{sec:method}.

Note, however, that the procedure outlined in section \ref{subsec:proc} prioritises higher tiers so that the cut ordering is already optimised and thus this parametric reasoning should not generally be necessary. Nevertheless, it might prove beneficial to the procedure in extreme cases, and indeed is applicable in other cutting techniques (such as the brute-force verifications of section \ref{subsec:complexity}).

\section{Generating random pseudo-structured circuits}
\label{app:gen-rand-circs}

The procedure presented in this paper, by design, works best on circuits that are highly structured. This raised an interesting problem when benchmarking, as testing on wholly random circuits would not properly showcase its effectiveness and, conversely, demonstrating only ideal example cases would not yield particularly informative results as such circuits could be made arbitrarily ideal, sending $\alpha\rightarrow0$. Consequently, for the benchmarking experiments, we generated random circuits in such a way as to include some localised structural elements, so as to avoid trivial (unstructured) cases, while also not simply designing ``best case'' circuits on which to experiment.

Specifically, for the main benchmarking tests (figure \ref{fig:results}), we generated circuits from random combinations of randomly placed T-CNOT-T sandwiches, Toffoli gates, CNOTs, and phase gates. We varied the number of such instances of these components in order to vary the T-counts of the circuits. Meanwhile, the circuits for the initial, small-scale verification experiments (section \ref{subsec:complexity}) were generated likewise, minus the Toffoli gates (as even one or two of these produce circuits too large to verify). Note that in randomly placing multi-qubit gates, each of its qubits may be placed randomly among the circuit's qubits, and note also that we express Toffoli gates in ZX-diagram form as follows:

\ctikzfig{toffoliZX}

This approach ensures we create appropriately random circuits of non-trivial local structure, akin to that of section \ref{subsec:tiered} or the composed-Toffolis circuits seen in \cite{Kissinger2020reducing}.

\section{Estimating efficiency}
\label{app:estimate}

Following the procedure of section \ref{subsec:proc} will result in a list of scalar terms whose sum gives the overall scalar value corresponding to the original Clifford+T circuit. However, for the purposes of benchmarking the effectiveness of the procedure, it is not important what that scalar value is, but rather how many scalar terms were produced (and hence the total runtime). With this in mind, one can determine an approximate upper-bound estimate of the number of terms that would be produced, in linear time, without having to compute all such terms.

This can be done by parameterising the branches of each cut into a single parametric graph (rather than two sets of distinct enumerated graphs), akin to \cite{sutcliffeParamZX} (and utilising the corresponding modified PyZX package of \cite{github:paramzx}). As such, every cut can be recorded parametrically on a single graph, without the need to append a list of distinct graphs, similar to the figures shown in appendix \ref{app:para-analysis}. By continuing the procedure on this parametric graph, one will ultimately arrive at a final Clifford graph of $n$ parameters (given $n$ cuts), which represents $2^n$ distinct graphs whose sum is the solution. Thus, the number of terms will be known to be $2^n$. This is, however, an upper-bound estimate as the parameterisation of the cuts can prevent much inter-step simplification that could be facilitated with exact numerical phases.

\end{document}